\documentstyle[aasms4,epsf]{article}

\begin{document}
\def\rpcomm#1{{\bf COMMENT by RP:  #1} \message{#1}}
\def\uptilde{\mathaccent"164}
\def\etal{{\it et al.~}}

\title{A counterrotating central component \\in the barred
galaxy NGC~5728} 

\author{F. Prada$^{1}$ and C. M. Guti\'errez$^{2}$}

\affil{$^{1}$Instituto de Astronom\'\i a, UNAM, 22830 Ensenada, Mexico.}
\affil{$^{2}$Instituto de Astrof\'\i sica de Canarias, 38200 La Laguna,
 Tenerife, Spain.}
 
\journalid{Vol}{Journ. Date}
\articleid{start page}{end page}
\paperid{manuscript id}
\cpright{type}{year}
\ccc{code}
\lefthead{Prada et al.}
\righthead{A counterrotating central component in NGC~5728}

\begin{abstract}

We present a detailed study of the stellar kinematics in the barred
galaxy NGC~5728 based on I-band photometry and long-slit spectroscopic observations in the region of the near-IR Ca~II triplet. The analysis of the stellar line-of-sight velocity distribution (LOSVD) has revealed, in the  central regions of the bar, the presence of a cold (v/$\sigma \sim$2.5)
prograde $S$-shaped velocity component that coexists in the central 4 kpc 
with a fainter and hotter (v/$\sigma$$\sim$0.5) counterrotating component. 
Beyond 4 kpc from the nucleus the LOSVD shows the stellar bar kinematics. 
The comparison of the radial surface brightness 
profile of the velocity components with that obtained from an 
I-band image shows that the counterrotating core follows a
r$^{1/4}$ profile, while the $S$-shaped component does not follows
the flat bar surface brightness profile. Several possible scenarios accounting for such kinematic signatures found in the center of the bar in NGC~5728 are
discussed.

 The data presented in this paper shows for the first time the presence
of extended retrograde motions in barred systems which, together with 
previous discoveries seems to indicate that the stellar counterrotation is
a phenomenon present all along the Hubble sequence.

\end{abstract}


\section{INTRODUCTION}

It has been reported in the last few  years the discovery of
counterrotating cores in elliptical galaxies (Franx \& Illingworth
1988; Bender et al. 1994; Rix \& White 1992) and the
presence of extended retrograde
motions  in the discs of early type spirals (Rix et al. 1992; 
Merrifield \& Kuijken 1994; Bertola et al. 1996). This phenomenon 
is  also present in the Sb galaxy
NGC~7331, where the bulge rotates opposite  to the disc (Prada et
al. 1996). The origin of the stellar counterrotation has been
extensively studied and generally attributed to the accretion of a
small satellite galaxy or  gradual infall of gas in a retrograde
orbit (e.g. Merrifield \& Kuijken 1994; Thakar \& Ryden 1996).
However, other alternatives are possible in terms of dynamical
instabilities (e.g. Wozniak \& Pfenniger 1997), that make it 
still unclear whether
the causes of counterrotation are of external or internal origin. Up to now no
observational evidence exists of extended stellar counterrotation in barred
galaxies, whose discovery would represent a
challenge to understand the significance of counterrotation throughout
the Hubble sequence. Having this in mind, we decided to conduct a
photometric and spectroscospic study of the SAB(r)a galaxy NGC 5728
(de Vaucouleurs et al. 1991). As a result of this
analysis it will be shown here that extended retrograde motions
can also be present in barred systems. In combination with previous
observations, our results demonstrate that counterrotation is present
all along the Hubble sequence; the study of this phenomenon could thus sheds
light to understand the mechanisms and the evolution of galaxy formation.

The paper is organized as follows: Section 2 gives details of the
observations and data reduction. Section 3 presents a brief description
of the structure of the galaxy. Section 4 describes the analysis of the
stellar line-of-sight velocity distribution (LOSVD).  Section 5 discuss 
on the nature of the kinematical components present in this
galaxy, and finally Section 6 gives the discussions and conclusions
of this work.

\section{OBSERVATIONS AND DATA REDUCTION}

Long-slit spectroscopic observations of NGC~5728 were performed on
several runs with the double arm ISIS spectrograph on the 4.2m WHT at
La Palma. In July 1996, the slit, 1.3$''$ wide, was placed along p.a.
30$^{\circ}$, close to the  bar major-axis (p.a. 33$^{\circ}$,
Schommer et al. 1988). We
used the red arm of ISIS with a 1200 $l/$mm grating centered at 8750
\AA\, including the near IR CaII triplet. The
dispersion was 0.38 \AA$\cdot$pixel$^{-1}$. Two
consecutive 1800 s exposures were recorded. As velocity template we
observed the KIII HR 5826 star. The seeing was  1.4$''$. A similar setup 
was used
in a previous run (February 1995), this time a 1$''$ slit wide was
placed along p.a.  86$^{\circ}$, and 23$^{\circ}$. A 600
l$/$mm grating was used giving a dispersion of 0.79 \AA$\cdot$pixel$^{-1}$.
 Two 1800 s frames were taken centered at 8750
\AA\ for p.a. 86$^{\circ}$, and one 1800 s exposure for p.a.
23$^{\circ}$. The G8 giant HR 7753 star was
observed as velocity template in this run.
The seeing was 
0.9$''$.  In both runs the slit length was 4$'$ and centered on the apparent
optical nucleus of NGC~5728. The detector was a $1024\times1024$ TEK CCD 
giving 0.36$''$ pixel$^{-1}$. The debias, flatfielding, 
wavelength calibration and sky subtraction was done in
the standard way using the IRAF package.

An I-band CCD image of NGC~5728 was taken with the BroCam CCD camera
on the 2.5m NOT at La Palma for the analysis of the morphology and 
comparison with the stellar kinematics. These observations were
done in August 1996 with a seeing of 0.7$''$. The camera
uses a $1024\times1024$ TEK CCD with 0.176 $''\cdot$pixel$^{-1}$. In 
total, we recorded a 40 minutes exposure. Figure 1 (Plate 1)
presents this image; the bar major-axis and the nuclear features
inside the spheroidal component of the galaxy are indicated. 
Hereafter, we will refer to p.a. 30$^{\circ}$ as the  bar major-axis.

\section{THE MORPHOLOGY OF NGC~5728}

An isophotal analysis of the I-band image of NGC 5728 was done using
Galphot (J\"orgensen, Franx, \& Kjaergaard 1992) to determine the
surface brightness, ellipticity, position angle and the C4 Fourier
coefficient along the semi-major axis radius (see Figure 2). The
structure of the galaxy can be summarized as follows: a nuclear
feature that appears elongated along p.a. 86$^{\circ}$ with a
semi-major axis radius of 3.5$''$ and a maximum ellipticity of 0.5;
the  stellar bar with a p.a. 33$^{\circ}$, a length of $\sim$60$''$,
maximum ellipticity of 0.7. Beyond the end of the  bar, the
ellipticity drops to 0.55 and the disk dominates. A star forming 
ring is located at $\sim$4$''$ (see Schommer et al. 1988; Wilson et al. 1993).
Interior to this ring the HST images by Wilson
et al.  (1993) reveal the nuclear feature as the presence of
bar-like continuum knots (called ABCD, named from East to West, 
beeing A and B very blue). This optical continuum light
within the ring can be interpreted in terms of a
nuclear stellar bar (Shaw et al. 1993; Wozniak et al. 1995), as star
formation triggered by the radio ejecta or as scattered light from the
active nucleus (see Wilson et al. 1993 for a discussion). The nature of
this nuclear feature will be discussed in the following sections.

In the central $\sim$20$''$, we have  a spheroidal component  
misaligned with respect to the bar major-axis (see Figure 1).  This is
also shown in the ellipticity profile at $\sim$10$''$ which has a
plateau of $\epsilon$$\sim$0.3 and an isophotal twist of
$\sim$8$^{\circ}$.

\section{THE STELLAR LINE-OF-SIGHT VELOCITY DISTRIBUTION IN NGC~5728}
 
The near-IR CaII triplet absorption features were used to determine 
the stellar LOSVD. The analysis was performed by means of the two-dimensional
unresolved Gaussian decomposition algorithm described in Prada et al.
(1996), and Prada, Guti\'errez, \& McKeith (1998). The resolution used
for the observations presented here, is 2 pixels and 1.4 pixels in
the spectral and spatial axes respectively  for p.a.  23$^{\circ}$ and
p.a. 86$^{\circ}$, and 3 pixels and 2.2 pixels respectively for the
bar major-axis spectra. Figure 3 shows on the left the spectra in
several positions along the  bar major-axis in the spectral region
analyzed, i.e. 8450-8700 \AA, together with the modeled spectra,
obtained by convolution of the derived LOSVD with the spectrum of the 
velocity template. The corresponding LOSVD are also plotted on the 
righthand side of Figure 3. 1-sigma error-bars for each LOSVD were
obtained by Monte Carlo simulations in the following way: first, we 
consider that the difference between the real and the modeled spectra 
gives an estimation of the error-bar in each point of the real spectra. 
Then, we simulate spectra considering that the astronomical signal 
corresponds to the modeled spectra and the noise follows a 
Gaussian distribution with the errorbar mentioned above. We run again 
our algorithm for each simulation obtaining the distribution of the 
LOSVD and taking the dispersion of this distribution as an estimation 
of the uncertainty in each velocity bin of the LOSVD.  Finally, we average 
over all velocity bins at each position. These error-bars constitute 
the statistical uncertainty relative to the
fit between the original and the recovered spectra and don't take into
account any possible effect due for example to the selection of the
stellar templates or modeling the shape of the galaxy's continuum.

We would like to mention that CaII in emission
is seen in some Seyfert 1 galaxies phisically associated with the 
BLR, however, there is no strong indication, generally, that
the Ca$^{+}$ zone is within a separate dynamical entity such as a disk (see
Persson 1988). He also find that the physical conditions
are so extreme in the Ca$^{+}$ zone that the calcium emission cannot
be a common feature of the spectra of Seyfert 1 galaxies and it 
it would be quite rare to see it in a Seyfert 2. NGC~5728 is classified
as Seyfert 2 by Philips et al. (1983), therefore we consider that the CaII 
in emission does not affect our results at all.

The LOSVD in the central $20''$ along the bar major-axis shows two 
distinct components with their velocity means at opposite sides
of the systematic velocity, except in the central 2$''$ where
the two components are not present. Beyond a radius of $\sim$12$''$ at 
either side of the nucleus the LOSVD  consists of a single component. The 
counterrotation is also confirmed along p.a. 23$^\circ$ where the
better seeing during this run allowed to see the two components in the LOSVD 
in the central 2$''$ (see Figure 4). While the LOSVD at p.a. 23$^\circ$
and the bar major axis are very similar in the central positions, there 
are some differences at distances $\sim 5-7$ arcsecs from the center. 
Although the position angles are close we believe that at least 
partially this could be a consequence of the kinematics
of the elongated physical structures associated with the barred component.

We have parameterized the LOSVD profiles by fitting one, and where
possible two Gaussians; this is enough to describe the major
kinematical features. Figure 5 shows the results of this analysis
plotting the radial velocities as a function of the radius along the bar
major-axis, p.a. 23$^{\circ}$ and p.a.  86${^\circ}$.
These can be described as a $S$-shaped velocity profile in the central
20$''$ along the  bar major-axis with $\sigma$$\sim$70$\pm$10 km
s$^{-1}$ and a maximum velocity of $\sim$170 km s$^{-1}$  at
$\pm$5$''$. Superimposed on this cold component is a fainter
counterrotating component that extends from 2$''$ out to 11$''$ at 
either side of the nucleus along the bar major-axis with 
a $\sigma$=95$\pm$10 km s$^{-1}$, and a slow rotation reaching 
$\sim$90 km s$^{-1}$ at $\pm$5$''$. In the central 2$''$,  a single 
Gaussian was fitted to the LOSVD with $\sigma$=165$\pm$15 km s$^{-1}$. 
A similar velocity pattern is also present along p.a. 23$^{\circ}$. At 
this position angle we fit two Gaussians to the LOSVD in central 2$''$ 
(see Figure 5). Beyond $\sim$20$''$ radius the velocity 
linearly rises with radius and the velocity dispersion is 
55$\pm$10 km s$^{-1}$. The kinematics in this region can be attributed 
to the stellar bar (see next section).

The velocities along p.a. 86$^\circ$ basically describe the overall
kinematics seen along the bar major-axis, only in the central 4$''$ the
LOSVD appears asymmetric. This might suggest the presence of the
counterrotating component seen at this position angle, since the fit of
a single Gaussian of $\sigma$$\sim$160$\pm$15 km s$^{-1}$ to the LOSVD
agrees with the velocity dispersion of a single Gaussian fit to the
LOSVD in the very center along the bar major-axis. However, higher spectral
resolution data is needed to confirm the presence of the two components
along p.a. 86$^\circ$.

In summary, we have found in the central parts of the 11 kpc (we adopt
a scale of 1$''\sim$ 190pc for H$\rm_o$=75 km s$^{-1}$Mpc$^{-1}$) bar of
NGC~5728  a cold (v/$\sigma$$\sim$2.5) prograde $S$-shaped velocity
component which coexist in the central 4 kpc with a dynamically hot
(v/$\sigma$$\sim$0.5) counterrotating component.
 
\section{THE NATURE OF THE CENTRAL BAR KINEMATICS}

A comparison between the photometric and the kinematic components
found in the center of the bar in NGC~5728, should shed some light onto the
nature and dynamics of such structures. For this purpose, we have compared the
surface brightness profile for a cut along p.a. 30$^{\circ}$ obtained
from the I-band image with the relative fluxes of the
velocity components found in our spectroscopic analysis along this 
position (see Figure 6).

On the basis of its kinematical and morphological properties we propose
to fit the surface brightness profiles of the three velocity components
inside 70$''$ with three components.  First, we model the bar by a flat bar profile ($I=I_o/(1+\exp((r-b)/a))$, which
describes the bar major-axis surface brightness profiles of early type
barred galaxies (Elmegreen et al. 1996). The parameters
found for NGC~5728 yield a characteristic scale-length of $b=53''$,
$a=4.1''$ for the region between 30$''$ and 60$''$. For the 
counterrotating component we propose an
$r^{1/4}$ law for the fit due to the
fact that this is a hot slowly rotating component and can be associated
with the spheroidal component present in the central 20$''$ of the
galaxy (see section 3). We get a very good fit with an effective radius
of 21.8$''$ for the region between 1$''$ and 9$''$. Finally, the 
brighter component in this region, i.e. the
$S$-shaped velocity profile has been fitted by an exponential profile of
4$''$ scale-length.  The sum of
the three modeled surface brightness profiles agrees very well with
the overall radial surface brightness I-band image cut along the  bar
major-axis (see Figure 6).

The question now is whether the counterrotating core and/or the
$S$-shaped velocity profile represent the kinematics of the bar in the
central regions. The $S$-shaped velocity component has the same
v/$\sigma$ as the bar kinematics outside a radius of 20$''$, but its
surface brightness profile does not follow the flat bar surface
brightness profile. The reason for this might be the contamination by
the nuclear feature and the star-forming ring inside the 5$''$
radius.  Following the discussion by Wilson et al. (1993), we agree
with them that the nuclear feature correspond to a blue 
component that dominates the light in the
very central regions (see Figure 5), but does not
contribute to the gravitational potential as indicated by its kinematics
(see Section 4). The reason why we state that the $S$-shaped component  is
associated to the barred kinematics is that the gaseous velocity field
in NGC~5728 (see Rubin 1980; Schommer et al. 1988) shows also an
$S$-shaped feature and follows the stellar kinematics at radii larger
than 20$''$.  The gaseous kinematics has been interpreted by
Schommer et al. (1988) in terms of  models of gas motions in
non-axisymmetric potentials and they claim that the star-forming ring
and the $S$-shaped velocity field may be related to an inner Lindblad
resonance. The gaseous kinematics also shows an expansion of $\sim$400
km s$^{-1}$ in the central 3.5$''$ but it does not show any
counterrotating signature.

 However, the $S$-shaped velocity profile can also be interpreted as
a very fast rotation due to a heavy mass concentration in the center of 
the bar in NGC~5728.  From the measured velocities in the very center, we 
can estimate an mass of $\sim$4 10$^{9}$ M$_{\odot}$ inside a radius of 
300 pc for an inclination of 48$^{\circ}$ (Schommer et al. 1988). This 
mass could be related to its Seyfert nucleus.

\section{DISCUSSION AND CONCLUSIONS}

Several scenarios have been proposed to explain kinematical structures
similar to the ones found in NGC~5728: dynamical instabilities,  the
accretion of a satellite or gas infall
with retrograde orbits. In the first case a non-axisymmetric structure
such as a bar modifies the stellar orbits, introducing retrograde
motions in its center, as shown in numerical simulations by
Wozniak \& Pfenniger (1997). In their models, the presence of 
retrograde orbits (almost
circular and perpendicular to the major-axis of the bar) inside the
corotation radius, can explain local counterrotation or even could
support a counterrotating core. One can expect that, if the mass on
such orbits increases, a critical mass ratio can be reached beyond
which the dynamics of the counterrotating component are decoupled from 
the direct bar. In
this scenario the presence of the counterrotating core in NGC~5728
could be interpreted as a secondary retrograde bar, and its presence
could thus be intrinsic to the primary bar. Indeed, n-body simulations
by Friedli (1996) have shown that a counterrotating secondary stellar
bar can be stable.

 However, we cannot rule out an external origin for the presence of 
retrograde motions in the center of NGC~5728 as shown by  
the  models of Thakar \& Ryden (1996). They explain 
the  counterrotation of the disks of spirals 
by the accretion of small satellites or gas with negative angular
momentum.

We conclude that the data presented in this paper reveal for the first
time the presence of extended retrograde motions in barred galaxies
which, together with previous discoveries in ellipticals, early-type
spirals and Sb galaxies (see Bertola \& Corsini 1997 for a review), 
seems to indicate
that counterrotation is a phenomenon present all along the
Hubble sequence. The scenarios mentioned above could explain
these, or alternatively a more general physical mechanism that allows
for galaxy transformation processes along the Hubble sequence might be needed.
In any case, the relevance of the phenomenon is clear and requires for a deeper
understanding, more statistics and two-dimensional spectroscopy.

\acknowledgements This research is based on observations obtained at
the WHT, operated by the ING and the NOT at the Observatorio del Roque
de los Muchachos (La Palma, Spain) of the IAC. We want to thank A.
Manchado for taking the NOT image and E. P\'erez, H. Zhao and the
referee for useful comments on the manuscript.

\newpage
 
\subsection*{FIGURE CAPTIONS}

\figcaption{An I-band grey-scale image of NGC 5728. The NE-SW 
and E-W arrows indicate the  bar major-axis  and the nuclear 
features at p.a. 86$^\circ$, respectively. Notice the misalignment 
of the spheroidal component with respect to the stellar bar.}

\figcaption{Isophotal analysis of the I-band surface photometry of NGC
5728. From top to bottom:  the major axis surface brightness profile,
the ellipticity profile, the position angle profile, and the C4 Fourier 
coefficient profile, plotted against the square root of the radius in
arcsec.}

\figcaption{({\it Left:}) The original and modeled spectra in the
region of the near-IR CaII triplet at different positions along the 
bar major-axis in NGC~5728. ({\it Right:}) The corresponding LOSVD. 
1-sigma error-bars for each LOSVD are shown (see text). SW is at top.}

\figcaption{A comparison of the LOSVD along
the bar major-axis and p.a. 23$^{\circ}$. The LOSVD along p.a. 86$^{\circ}$
are also shown; West is at top (see text). Error-bars as in Fig. 3.}

\figcaption{ Radial velocities determined by fitting two Gaussians to
the LOSVD  along the  bar major-axis ({\it top}),
p.a.  23$^{\circ}$ ({\it middle}), and p.a.  86$^{\circ}$ ({\it
bottom}). Open and filled squares refer to the fainter and brighter
components present in the LOSVD. The triangles
refer to a single component in the LOSVD. The open triangles in 
p.a.  86$^{\circ}$ refer to the fit of a single Gaussian to the 
asymmetric LOSVD in the central 3.5$''$. The dashed line correspond to the continuum at $\sim$8500 \AA\ obtained from the spectroscopic data along p.a. 86$^{\circ}$. The letters ABD indicate the 
position of the features named by Wilson et al. (1993) in the HST
continuum images of NGC~5728.}

\figcaption{The surface brightness profile  of the three
velocity components obtained from the LOSVD along
the  bar major-axis, i.e. the counterrotating core (open squares),
the cold $S$-shaped component (filled squares), and the bar
(triangles). The open circles represent the sum of the 
counterrotating and $S$-shaped surface brightness profiles. Also 
plotted is the surface brightness 
cut along the bar major-axis  obtained from the I-band image (solid line). 
The dotted lines are the modeled surface brightness profiles for the 
three velocity
components. The dashed line is the sum of these three model components.
(see text).}

\newpage

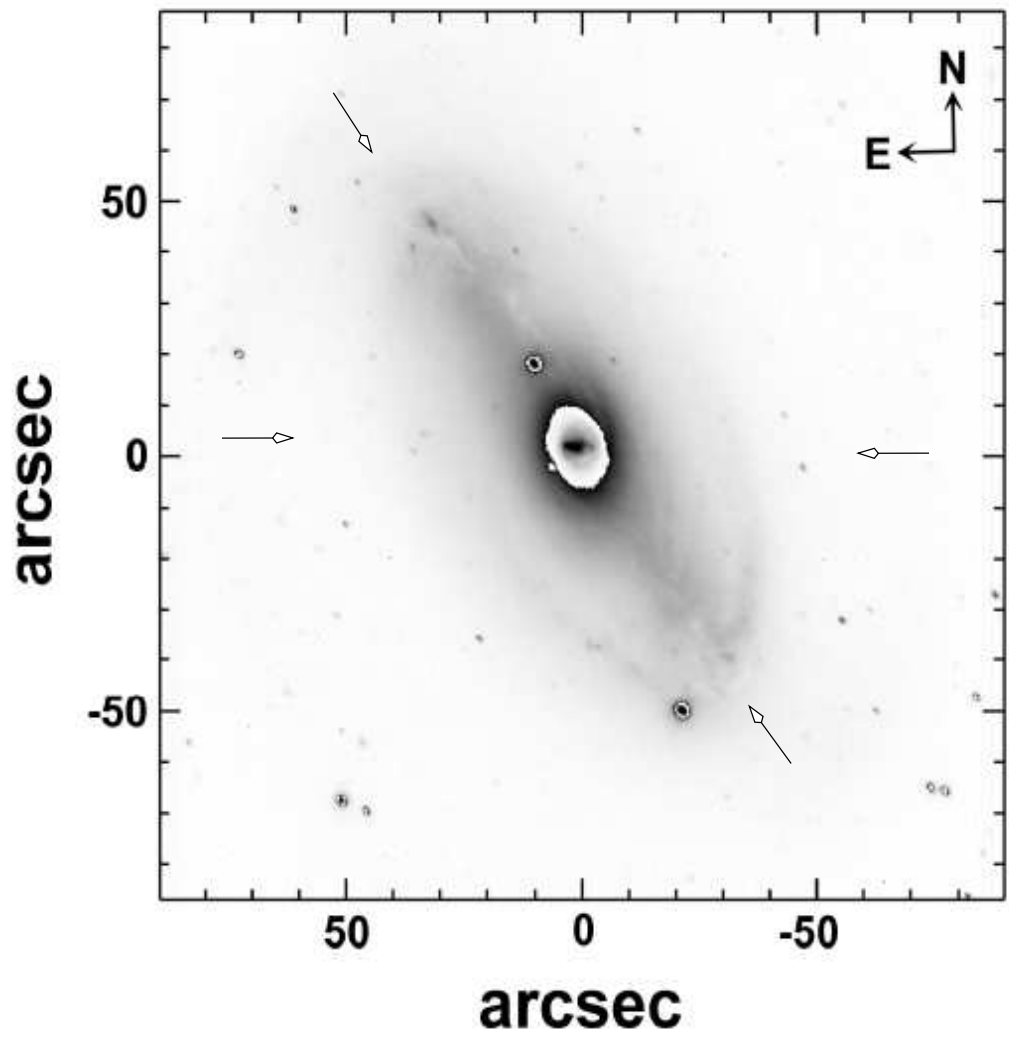
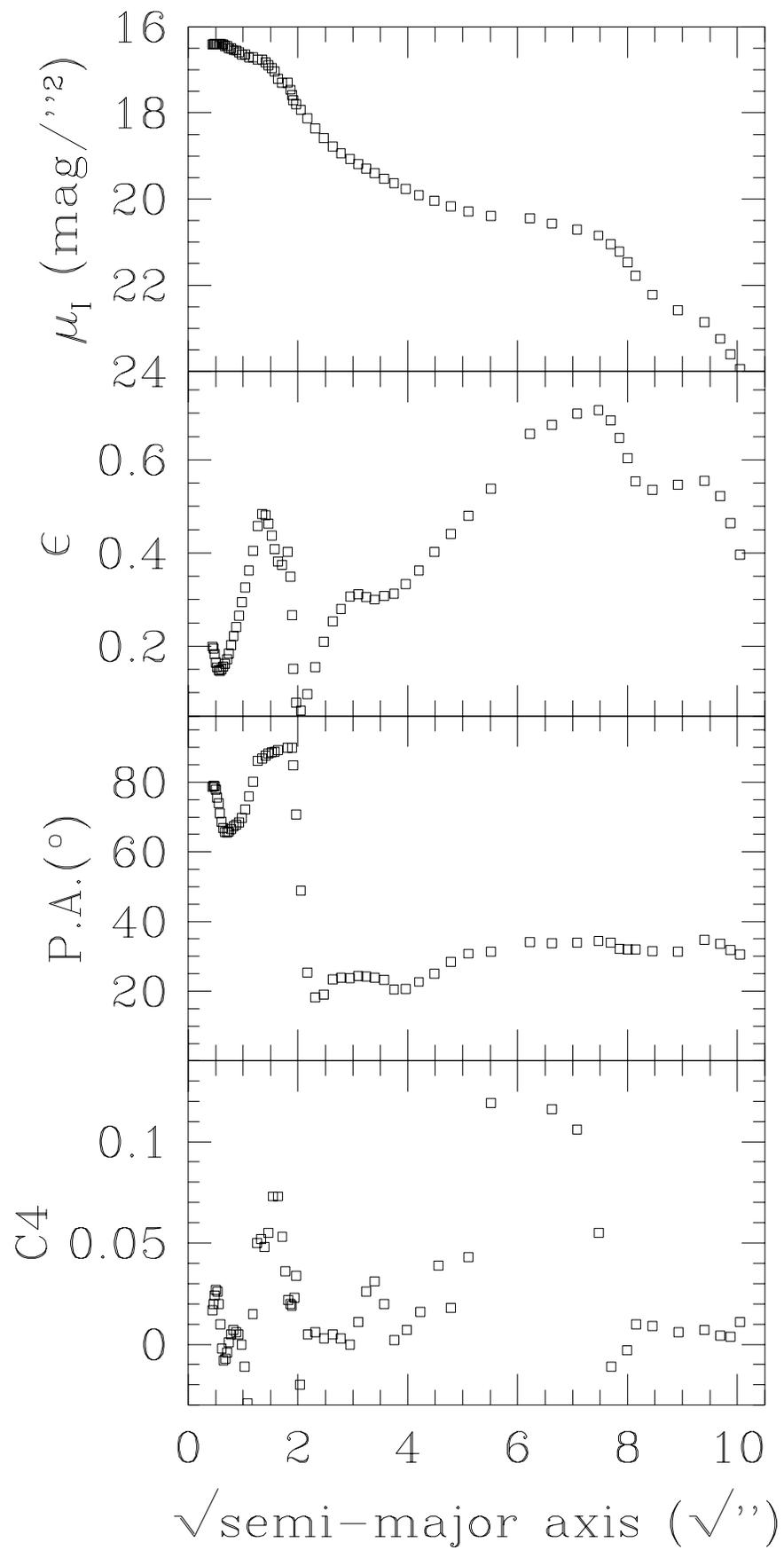
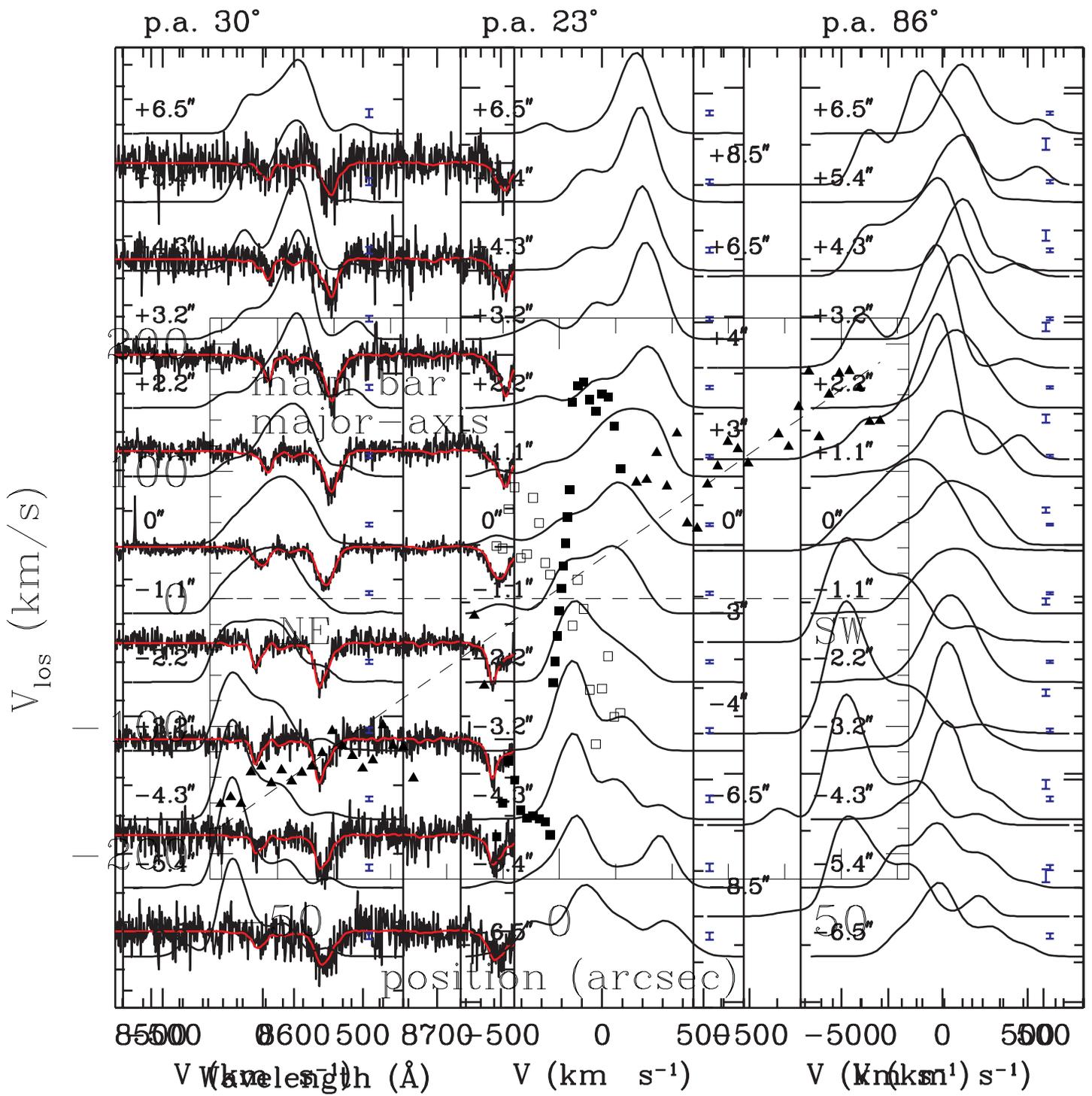
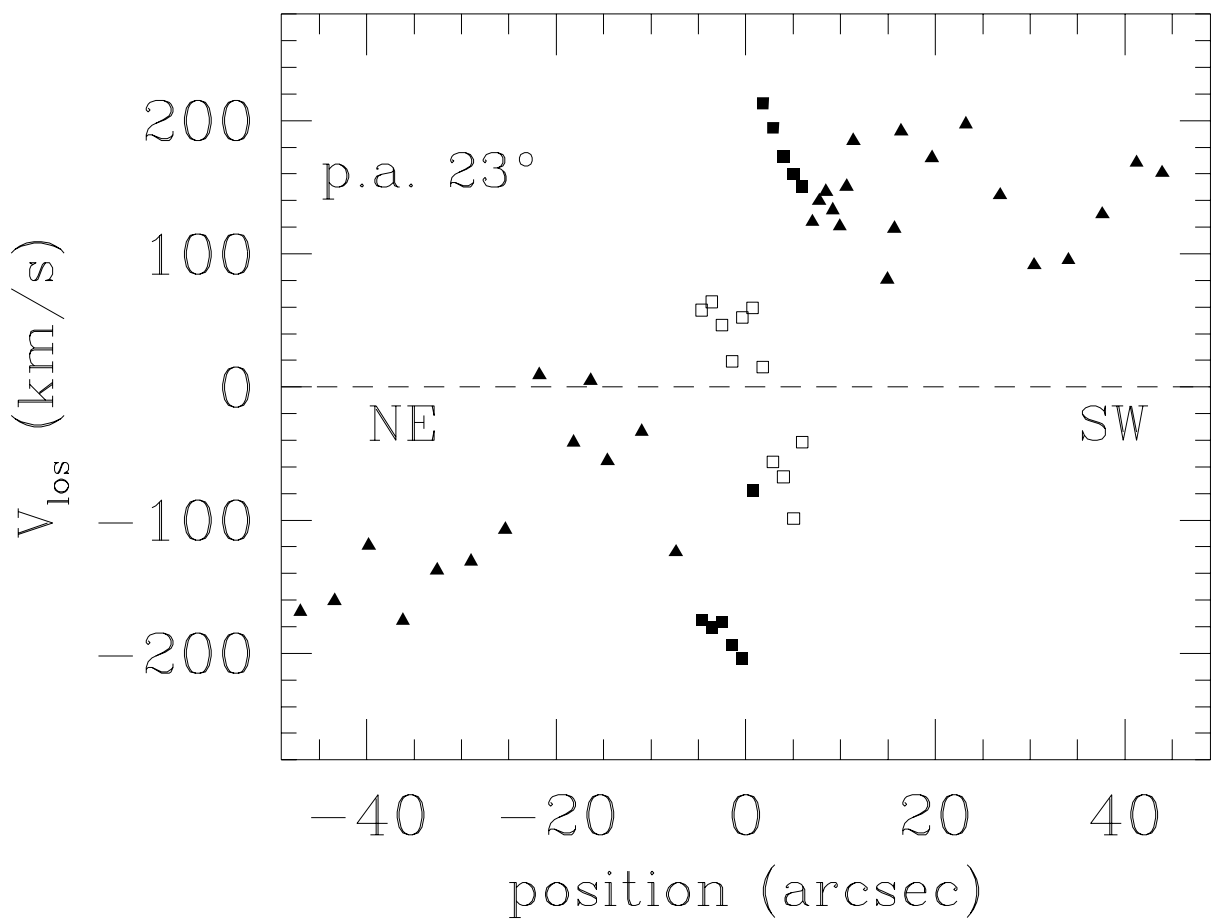
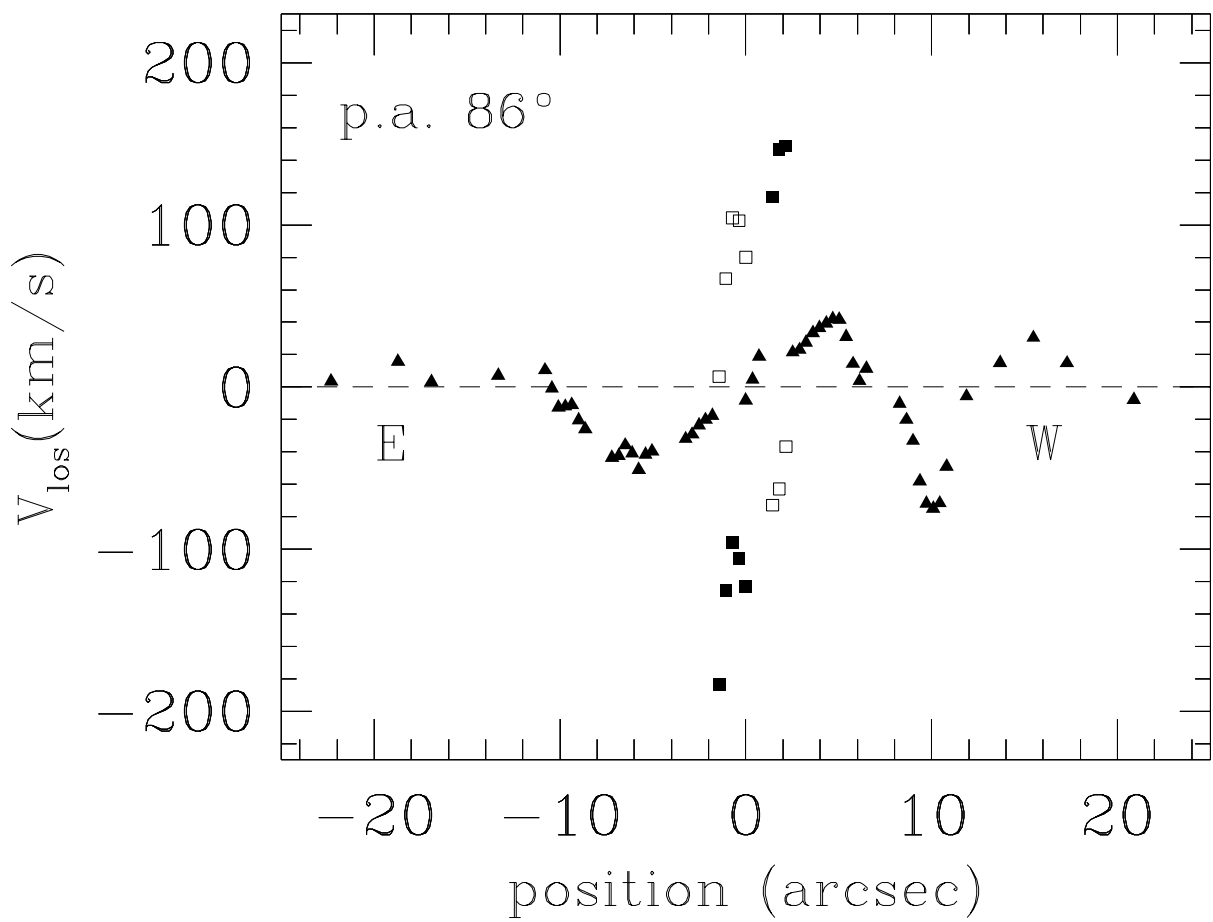
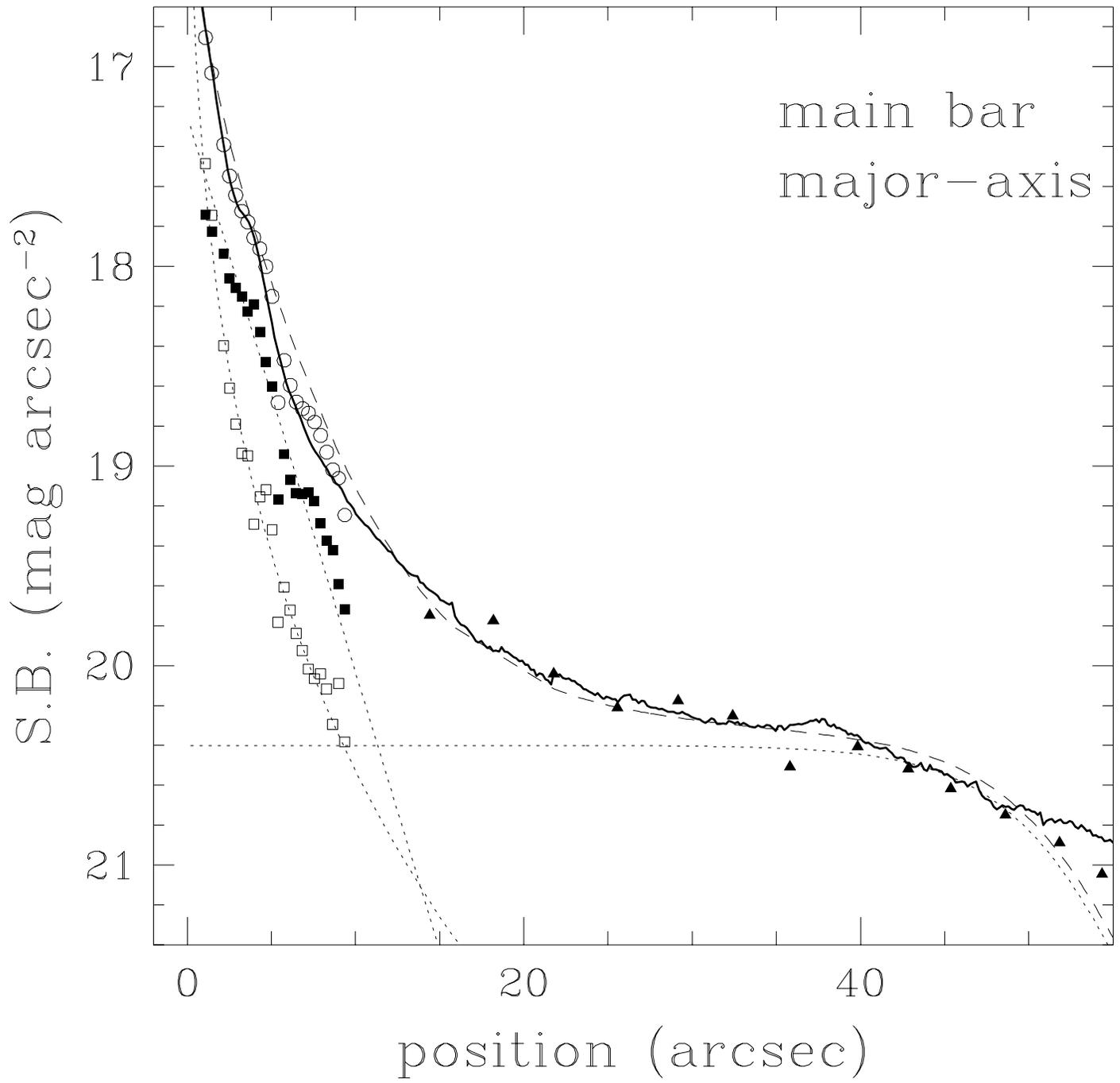
\end{document}